\documentclass[useAMS,usenatbib]{mn2e}
\usepackage{graphicx,amsmath}
\usepackage[usenames]{color}
\usepackage{aas_macros}
\title[Pulsation frequency distribution in $\delta$~Scuti stars]
{Pulsation frequency distribution in $\delta$~Scuti stars}
\author[L.A. Balona and J. Daszy\'nska-Daszkiewicz] 	
{L.A. Balona$^1$, J. Daszy\'nska-Daszkiewicz$^2$, A.A. Pamyatnykh$^3$\\
\\
$^1$South African Astronomical Observatory, P.O. Box 9, Observatory, Cape
Town, South Africa\\
$^3$Instytut Astronomiczny, Uniwersytet Wroc{\l}awski, Kopernika 11, 51-622
Wroc{\l}aw, Poland\\
$^3$Copernicus Astronomical Center, Bartycka 18, 00-716 Warsaw, Poland
}

\begin{document}

\date{Accepted .... Received ...}

\pagerange{\pageref{firstpage}--\pageref{lastpage}} \pubyear{2011}

\maketitle

\label{firstpage}

\begin{abstract}
We study the frequency distributions of $\delta$~Scuti stars observed
by the {\it Kepler} satellite in short-cadence mode.  To minimize errors in
the estimated stellar parameters, we divided the instability strip into ten
regions and determined the mean frequency distribution in each region.  We
confirm that the presence of low frequencies is a property of all
$\delta$~Scuti stars, rendering meaningless the concept of
$\delta$~Sct/$\gamma$~Dor hybrids. We obtained the true distribution of
equatorial rotational velocities in each region and calculated the frequency
distributions predicted by pulsation models, taking into account rotational
splitting of the frequencies. We confirm that rotation cannot account for the
presence of low frequencies.  We calculated a large variety of standard
pulsation models with different metal and helium abundances, but were unable
to obtain unstable low-frequency modes driven by the $\kappa$~mechanism in 
any model.  We also constructed models with modified opacities in the envelope.
Increasing the opacity at a temperature $\log T = 5.06$ by a factor of two does 
lead to instability of low-degree modes at low frequencies, but also decreases 
the frequency range of $\delta$~Sct-type pulsations to some extent.
We also re-affirm the fact that less than half of the stars in the
$\delta$~Sct instability strip have pulsations detectable by {\it Kepler}.
We also point out the huge variety of frequency patterns in stars with roughly
similar parameters, suggesting that nonlinearity is an important factor in
$\delta$~Sct pulsations.
\end{abstract}


\section{Introduction}

Delta Scuti stars are main sequence dwarfs and giants with spectral types
A0--F5 which pulsate in multiple p, g and mixed modes.  They lie on the extension
of the Cepheid instability strip towards low luminosities. From ground-based
observations, most $\delta$~Sct stars are found to pulsate with frequencies in
the range 5--50\,d$^{-1}$.  The pulsations are driven by the $\kappa$ mechanism
due to partial ionization of He\,II.

The $\gamma$~Doradus stars lie in a fairly small region on, or just above, the
main sequence that partly overlaps the cool edge of the $\delta$~Sct
instability strip.  They pulsate in multiple  g modes with low frequencies
(typically less than 5\,d$^{-1}$) and are thought to be driven by a mechanism
known as ``convective blocking'' \citep{Guzik2000, Dupret2004a}.  A subset of
the $\delta$~Sct stars pulsate in both the high and low-frequency ranges and
are known as $\delta$~Sct/$\gamma$~Dor hybrids.

Space photometry from the {\it Kepler} observatory has completely changed
this view.  It is found that {\em all} $\delta$~Sct stars pulsate in
low-frequency modes, i.e. they are {\em all} hybrids, \citep{Balona2014a}.
The misconception of hybrid $\delta$~Sct/$\gamma$~Dor stars is due to the fact
that low frequencies in $\delta$~Sct stars have sufficiently high amplitudes
to be observed from the ground only in a narrow band close to the red edge of
the instability strip.  The low frequencies in $\delta$~Sct stars present a
serious problem because standard models predict that low frequencies are stable
for $\delta$~Sct stars hotter than the granulation boundary.  In these hot stars
the subsurface convection zone is too thin for convective blocking to drive
low-frequency pulsations.

Apart from the problem of low frequencies in hot $\delta$~Sct stars, there
is the additional problem that the majority of stars in the $\delta$~Sct
instability strip do not pulsate \citep{Balona2011g}.  It is difficult to
understand why in two stars with the same physical parameters one star should
pulsate as a $\delta$~Sct star and the other not pulsate at all, or at least
have pulsational amplitudes not detectable in {\it Kepler} observations.
It is possible that nonlinear mode coupling may stabilize the pulsations
\citep{Dziembowski1982, Dziembowski1985a, Dziembowski1988} or that
the opacity driving mechanism may be saturated \citep{Nowakowski2005}.
Since nonlinear models of nonadiabatic nonradial pulsations do not exist, 
it is not possible, at present, to test these ideas.

Another issue is the fact that A stars appear to be not as simple as
previously thought.  It has always been supposed that stellar activity
ceases for stars hotter than the granulation boundary because the subsurface
convection zone is too thin to generate a magnetic field by the dynamo
mechanism.  Analysis of {\it Kepler} photometry shows that about 40\,per cent
of A stars have light variations whose periods closely match the expected
rotation periods of these stars, suggesting the presence of starspots 
\citep{Balona2011e, Balona2013c}.  Furthermore, about 2\,per cent of A stars 
have flares \citep{Balona2012c, Balona2013c}.  In fact, the relative number of
A-type flare stars is about the same as F and G flare stars and not much 
smaller than K and M flare stars \citep{Balona2015a}.  The starspot and
flare activity in A stars indicates that we do not fully understand the 
physics of stellar envelopes which may be related to our lack of understanding
of low frequencies in hot $\delta$~Sct stars.

Using {\it Kepler} data, \citet{Balona2014a} showed that low frequencies in
$\delta$~Sct stars are not stochastically driven, nor can they be explained as
nonlinear combinations of high-frequency modes.  By comparing the observed
distribution of frequencies in {\it Kepler} $\delta$~Sct stars with those from
models, \citet{Balona2014a} also showed that the low frequencies cannot be
explained as a result of rotational splitting of high-frequency modes.

In order to understand the origin of the low frequencies in $\delta$~Sct
stars, it is necessary to explore what effects contribute most to the
stability or instability of these frequencies in the models.  A simple 
examination of the instability parameter, $\eta$, calculated by a nonadiabatic
pulsation model will tell us if a particular mode is stable or unstable.   
However, in order to compare the predicted frequency distribution with 
observations, it is necessary to simulate rotational splitting of the 
frequencies as well as other effects.  In the absence of mode identification 
and sufficiently accurate stellar parameters, this is the only way to
test if model predictions agree with observations.  In order to simulate the 
effect of rotational splitting, the distribution of true equatorial rotational 
velocities for the group of stars under study is required.  In 
\citet{Balona2014a} all $\delta$~Sct stars were treated as a single group 
using rotation periods inferred from the light curve.

In this paper we compare the observed and simulated frequency distributions in
several regions across the instability strip.  Different physical conditions
are used for each set of models in an attempt to determine which effects
most closely reproduce the observed frequency distributions. In this way we
hope to determine whether it is possible to understand the low frequencies
under the assumption that they are driven by the $\kappa$~mechanism.  We 
study mode stability using standard models, i.e. using different abundances 
and standard opacities, as well as non-standard models, i.e. using opacities 
which are artificially increased in certain regions in the envelope.

\section{The data}

The {\it Kepler} light curves are available as uncorrected simple aperture
photometry (SAP) and with pre-search data conditioning (PDC) in which
instrumental effects are removed.  The vast majority of the stars are
observed in long-cadence (LC) mode with exposure times of about 30\,min.
Several thousand stars were also observed in short-cadence (SC) mode with
exposure times of about 1\,min.  These data are publicly available on the
Barbara A. Mikulski Archive for Space Telescopes (MAST, {\tt
archive.stsci.edu}).

In order to identify $\delta$~Sct stars in the {\it Kepler} field, we
calculated the periodograms of over 20\,000 stars, including all stars
observed in SC mode.  By visually examining the light curves and
periodograms and using the effective temperatures in the {\it Kepler}
Input Catalogue (KIC, \citealt{Brown2011a}) as a guide, we were able to
identify over 1600 $\delta$~Sct stars in LC mode and 403 $\delta$~Sct
stars in SC mode with known stellar parameters.  These do not include
$\delta$~Sct stars in eclipsing binary systems.

The highest pulsation frequency that can be detected in LC mode is about
24\,d$^{-1}$.  Since the pulsation frequencies in many $\delta$~Sct stars 
exceed this value, only SC observations allow the unambiguous identification 
of all pulsation frequencies.  Due to variations in heliocentric time
correction, the {\it Kepler} data are not sampled at exactly equal time
intervals.  In principle, this allows frequencies higher than 24\,d$^{-1}$
to be identified in LC data \citep{Murphy2013a}.  In practice, however,
the chance of frequency misidentification is very high for low-amplitude
peaks because the difference in peak amplitude between the high- and
low-frequency aliases is slight.  For this reason, we restrict our analysis
to the 403 $\delta$~Sct stars observed in SC mode.

To extract all significant frequencies we first calculated the standard
Lomb periodogram for unequally-spaced data \citep{Press1989}.  The usual
technique to extract peaks is that of successive prewhitening.  Great caution
needs to be exercised in using this method as indiscriminate use leads to a
large number of spurious frequencies \citep{Balona2014d}.  In effect, only
those peaks which are visible in the periodogram of the raw, un-prewhitened
data can be considered as significant.  Indiscriminate successive prewhitening 
will extract frequencies which cannot be resolved in the periodogram, even 
though the extracted amplitudes of such spurious frequencies may be deemed
significant.  In determining the frequency content of a star, we terminated
prewhitening when the peak can no longer be resolved in the original
periodogram.

\begin{figure}
\centering
\includegraphics{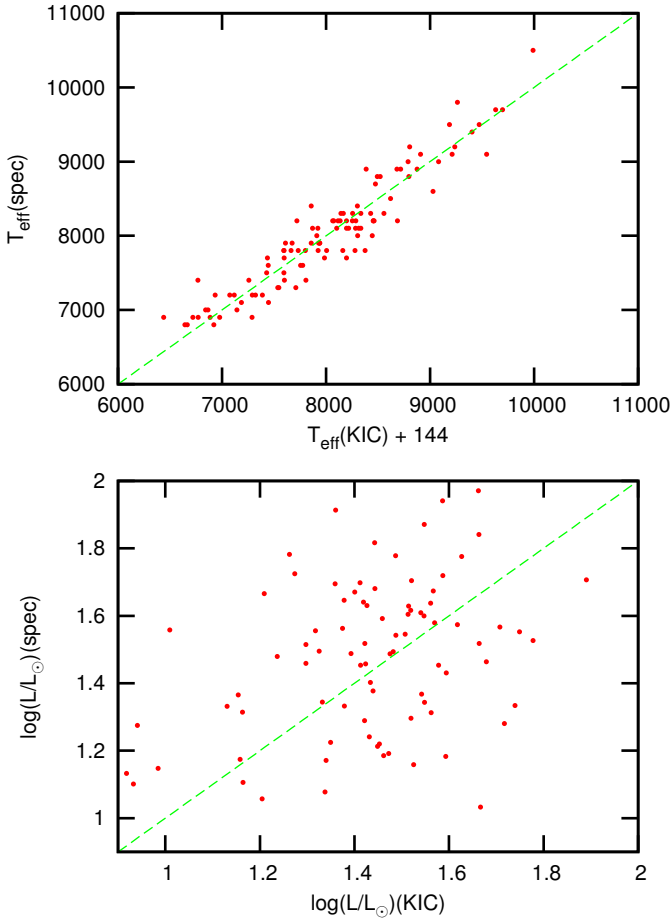}
\caption{Top panel: the corrected KIC effective temperature as a function of the
effective temperature determined by spectroscopic observations of
\citet{Niemczura2015}.  The straight line has unit slope and zero intercept.
Bottom panel: the value of $\log(L/L_\odot)$ derived from the KIC parameters
as a function of $\log(L/L_\odot)$ derived from the spectroscopic parameters.  The
straight line has unit slope and zero intercept.}
\label{spkic}
\end{figure}

\section{Systematic and random errors}

In order to compare the observed frequency distribution with the calculated
frequency distribution we need to determine the accuracy of the stellar
parameters for individual stars.  The stellar parameters derived from Sloan
multicolour photometry (without the u band) are listed in the KIC.  It is
important to determine the systematic and random errors that might be
present in the values of effective temperature, $T_{\rm eff}$, and relative
luminosity, $\log L/L_\odot$, derived from the KIC parameters.

We do not know the true values of $T_{\rm eff}$ and $\log L/L_\odot$ for any
$\delta$~Sct star, but we can compare the values in the KIC derived from
multicolour photometry with the more precise values that can be derived from
spectroscopy.  Recently, \citet{Lehmann2011, Catanzaro2010, Niemczura2015}
and \citet{Tkachenko2013b} have obtained spectroscopy of several {\it
Kepler} stars from which stellar parameters are derived.  The most numerous
observations are those of \citet{Niemczura2015} for which we find a systematic
difference in effective temperature $\Delta T_{\rm eff} = T_{\rm eff}({\rm spec}) -
T_{\rm eff}({\rm KIC}) = 144 \pm 24$\,K from 107 stars with a standard
deviation of 253\,K per star.  There is no significant dependence of
$\Delta T_{\rm eff}$ on $T_{\rm eff}$.  Since the standard deviation of
$T_{\rm eff}({\rm spec}) \approx 150$\,K \citep{Niemczura2015}, we estimate
that the standard deviation of $T_{\rm eff}({\rm KIC}) \approx 200$\,K on
the assumption that the variances add in quadrature.  The top panel of 
Fig.\,\ref{spkic} shows $T_{\rm eff}({\rm spec})$ from \citet{Niemczura2015} 
as a function of $T_{\rm eff}({\rm cor}) = T_{\rm eff}({\rm KIC}) + 144$.

Given $T_{\rm eff}$ and the surface gravity, $\log g$, the stellar radius may
be estimated using the relationship in \citet{Torres2010a}.  This
relationship requires the metal abundance which is usually listed among the
spectroscopically derived parameters.  When the metal abundance is not available, 
the solar value was used.  From the effective temperature and the radius, the 
relative stellar luminosity, $\log(L/L_\odot)({\rm spec})$, can be
determined.  For stars where only KIC values are available, the luminosity, 
$\log(L/L_\odot)({\rm KIC})$ can likewise be estimated from the tabulated 
effective temperature and radius.

The bottom panel of Fig.\,\ref{spkic} shows $\log(L/L_\odot)({\rm spec})$ as
a function of $\log(L/L_\odot)({\rm KIC})$.  The correlation is rather poor,
no doubt due to the poor precision of $\log(g)$ from the photometry.
According to \citet{Brown2011a} the standard deviation of $\log g$ in the KIC
for dwarfs is about 0.4\,dex, whereas it is about 0.15\,dex from the
spectroscopy \citep{Niemczura2015}.  As the figure shows, a line of unit slope
and zero intercept gives a satisfactory fit.  The standard deviation in
$\log(L/L_\odot)$ is 0.26 if the regression is performed with
$\log(L/L_\odot)({\rm KIC})$ as the independent variable or 0.29 as the
dependent variable.  The standard deviation of $\log(L/L_\odot)$ derived
from the KIC parameters must therefore be about 0.20\,dex, since the total 
variance is the quadratic sum of the two variances.

\begin{figure}
\centering
\includegraphics{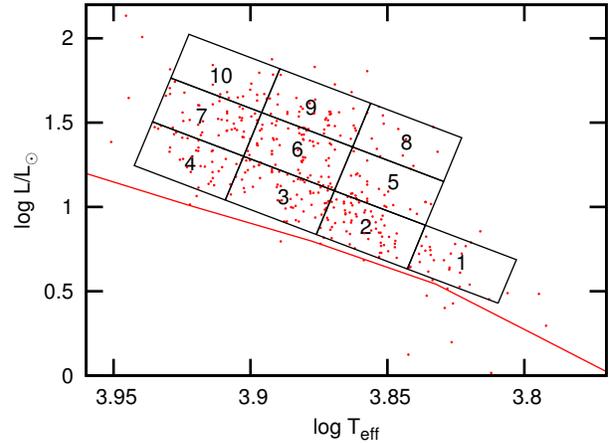}
\caption{The theoretical H-R diagram showing the SC $\delta$~Sct stars
(dots) and the designated regions (labeled).  The line is the theoretical
zero-age main sequence.}
\label{regions}
\end{figure}

We used spectroscopically derived parameters whenever possible for each star.
When only KIC values are known, we corrected $T_{\rm eff}$ listed by the
KIC by 144\,K and estimated the luminosity from the KIC radius and 
uncorrected effective temperature.

\begin{center}
\begin{table*}
\caption{For each region number, $N_{\rm Reg}$, the number of $\delta$~Sct
stars, $N$, in that region and the average number of modes, $N_{\rm modes}$, 
having $\nu > 5$\,d$^{-1}$ are shown.  Also shown is the corresponding range 
in spectral type, the mean effective temperature, $T_{\rm eff}$,  luminosity, 
$\log L/L_\odot$, surface gravity, $\log g$ and radius, $R/R_\odot$. The last 
column, $\langle v_e \rangle$ is the mean equatorial rotational velocity 
(in km\,s$^{-1}$) derived from observations of many field stars of the
particular spectral type and luminosity class range.}
\label{tab:reg}
\begin{tabular}{rrrlrrrrr}
\hline
\multicolumn{1}{c}{$N_{\rm Reg}$} &
\multicolumn{1}{c}{$N$}   &
\multicolumn{1}{c}{$N_{\rm modes}$}   &
\multicolumn{1}{c}{Sp.Ty.}   &
\multicolumn{1}{c}{$\log T_{\rm eff}$} &
\multicolumn{1}{c}{$\log L/L_\odot$} &
\multicolumn{1}{c}{$\log g$} &
\multicolumn{1}{c}{$R/R_\odot$} &
\multicolumn{1}{c}{$\langle v_e \rangle$} \\
\hline
  1 & 29 &  90 & F8V--F3V     & $3.8253 \pm 0.0015$ & $0.6889 \pm 0.0152$ & $4.100 \pm 0.011$ & $1.654 \pm 0.024$ &   40  \\
  2 & 60 & 252 & F4V--A9V     & $3.8546 \pm 0.0012$ & $0.8896 \pm 0.0119$ & $4.043 \pm 0.009$ & $1.822 \pm 0.020$ &   90  \\
  3 & 32 & 183 & F0V--A7V     & $3.8858 \pm 0.0019$ & $1.0922 \pm 0.0178$ & $4.003 \pm 0.011$ & $1.993 \pm 0.030$ &  130  \\
  4 & 25 & 138 & A8V--A4V     & $3.9146 \pm 0.0012$ & $1.2671 \pm 0.0190$ & $3.985 \pm 0.013$ & $2.136 \pm 0.039$ &  150  \\
  5 & 35 & 238 & F5IV--F1IV   & $3.8537 \pm 0.0014$ & $1.0808 \pm 0.0136$ & $3.869 \pm 0.010$ & $2.280 \pm 0.030$ &   70  \\
  6 & 61 & 245 & F1IV--A8IV   & $3.8837 \pm 0.0012$ & $1.3338 \pm 0.0130$ & $3.781 \pm 0.010$ & $2.663 \pm 0.035$ &  110  \\
  7 & 34 & 220 & A8IV--A5IV   & $3.9121 \pm 0.0014$ & $1.4897 \pm 0.0148$ & $3.777 \pm 0.013$ & $2.793 \pm 0.047$ &  140  \\
  8 & 18 & 103 & F5III-F1III  & $3.8471 \pm 0.0022$ & $1.3958 \pm 0.0210$ & $3.565 \pm 0.016$ & $3.381 \pm 0.070$ &   80  \\
  9 & 58 & 365 & F1III-A8III  & $3.8746 \pm 0.0013$ & $1.5508 \pm 0.0118$ & $3.551 \pm 0.009$ & $3.560 \pm 0.041$ &  100  \\
 10 & 15 &  83 & A9III-A6III  & $3.9043 \pm 0.0025$ & $1.7030 \pm 0.0181$ & $3.553 \pm 0.012$ & $3.692 \pm 0.059$ &  110  \\
\hline
\end{tabular}
\end{table*}
\end{center}

Given the large errors in $T_{\rm eff}$ and $\log L/L_\odot$, the lack
of mode identification, and the profound effect of rotation on the
frequency spectrum, it is not possible to model the pulsation frequencies
of individual stars.  In fact, we are not interested in detailed modeling
of a specific star, but only in a comparison of the overall observed and
calculated frequency distributions.  For this reason we have taken a different
approach.  We can minimize the uncertainties in  effective temperature and
luminosity by using the mean values of $\log T_{\rm eff}$ and $\log
L/L_\odot$ of many stars in approximately the same location in the instability
strip.  A good estimate of the mean frequency distribution at this position in
the instability strip is obtained by averaging the frequency distributions of
these stars.  For this purpose we divided the theoretical H-R diagram into ten
regions as shown in Fig.\,\ref{regions}.  These regions were chosen to cover
the instability strip as evenly as possible with reasonable resolution while,
at the same time, allowing a fairly large number of stars in each region.  The
mean effective temperature, luminosity, surface gravity and radius of all
$\delta$~Sct stars in each region are shown in Table\,\ref{tab:reg}.

\section{Mode visibility}

In determining the frequency distribution of a particular star we are faced
with the problem that individual frequencies are of widely different
amplitudes.  It would, of course, be preferable to calculate the amplitudes
from the models as well.  Unfortunately nonlinear calculations of nonradial
modes are beyond existing technical capabilities.  One approach is to take
the linear growth rate as a proxy for the amplitude.  However, the final
pulsational amplitude attained by a star depends on many factors and not
only the initial growth rate.  We decided that the most satisfactory
method is to ignore amplitude altogether. In other words, a mode of very
high amplitude is given the same weight as a mode of very low amplitude.  In
determining the observed frequency distribution, we simply count the number
of significant frequency peaks in the periodogram within a given frequency
interval, ignoring their amplitudes.

Using this approach requires that each unstable mode in the models
be weighted according to its visibility.  It is evident that modes of very
high spherical harmonic degree, $l$, will not be observed because of
cancellation effects and cannot therefore be given the same weight as modes
of low $l$.  In other words, each unstable mode derived from a model
should be assigned a weight equal to its visibility before it can be
compared with observations.  To calculate the visibility of a mode, we need to
discuss the expected relative light amplitude caused by a mode of particular
spherical harmonic degree, $l$,  and azimuthal order, $m$.

The monochromatic light amplitude, $A_\lambda(i)$, for a pulsating star
whose axis of rotation is inclined at angle $i$ with respect to the observer
can be expressed as \citep{DaszynskaDaszkiewicz2002}:
\begin{align*}
A_\lambda(i) &= \epsilon Y_l^m (i, 0) b_l^\lambda Re\{D_{1,l}^\lambda + D_{2,l} +
D_{3,l}^\lambda\}.
\end{align*}
Here $\epsilon$ is the intrinsic amplitude, $Y_l^m(i,\phi) = N_{lm}P_l^m(\cos i)e^{\mathrm{i}m\phi}$
is the spherical harmonic with $N_l^m$ the normalizing factor and
$P_l^m(\cos i)$ the associated Legendre polynomial ($\mathrm{i} = \sqrt{-1}$).  The disc
averaging factor, $b_l^\lambda$ can be calculated for any particular limb-darkening
law.  In the above expression, the $D_{1,l}$ term describes the effect of
temperature variation on the light amplitude,  whereas the influence of
effective gravity changes is contained in the $D_{3,l}$ term.  The $D_{2,l}$
term describes the effect of variations in projected area on the light
amplitude.  These terms are given by
\begin{align*}
D_{1,l}^\lambda &= \frac{1}{4} f \frac{\partial \log\mathcal{F}_\lambda |b_l^\lambda|}{\partial \log(T_{\rm eff})},\\
D_{2,l} &= (2 + l)(1 - l),\\
D_{3,l}^\lambda &= -\left(\frac{3\omega^2}{4\pi\bar{\rho}}+2\right)
 \frac{\partial \log\mathcal{F}_\lambda |b_l^\lambda|}{\partial \log(g_{\rm eff})}.
\end{align*}
The ratio of luminous flux to displacement, $f$, for each mode is calculated
by the nonadiabatic pulsation code.  The partial derivatives of the luminous flux, 
$\mathcal{F}_\lambda$, with respect to effective temperature, $T_{\rm eff}$, and 
with respect to effective gravity, $g_{\rm eff}$, can be obtained from model 
atmospheres.  The angular pulsation frequency is $\omega$, and $\bar{\rho}$ is 
the mean density of the star.

If we consider a group of stars in the same region, the visibility of a
particular mode for the group is obtained by averaging the various terms in
the above equation.  Because we are only counting unstable modes,
irrespective of amplitude, $\epsilon$ is unity if the mode is
unstable and zero if it is stable.  The average value of $Y_l^m(i,0)$ is
found assuming random orientation of the axis of rotation so that,
putting $x = \cos(i)$,
\begin{align*}
\langle Y_l^m (i, 0)\rangle &= N_l^m\langle |P_l^m(\cos i)|\rangle,\\
&= N_l^m\frac{2}{\pi}\int_0^1 \frac{|P_l^m(x)|}{\sqrt{1-x^2}} dx,
\end{align*}
with
\begin{align*}
N_l^m &= \sqrt{ \frac{(2l+1)(l-m)!}{4\pi (l+m)!} }.
\end{align*}
The value of $\langle Y_l^m (i, 0)\rangle$ does not change very much and
lies within the range 0.13--0.30 for modes with $l \le 6$.

The disk-averaging factor, $b_l^\lambda$, is given by
\begin{align*}
b_l^\lambda &= \int_0^1 h_\lambda (\mu)\mu P_l(\mu) d\mu,
\end{align*}
where $\mu = \cos\theta$ and $\theta$ is the angle in the spherical
coordinate system centered on the star.  We used the simple limb-darkening
law $h(\mu) = 1 + \frac{3}{2}\mu$ independent of wavelength.  Values of
$b_l$ drop sharply with $l$: $b_0 = 1, b_1 = 0.708, \dots b_6 = 0.008$.

In calculating $D_{1,l}$ and $D_{3,l}$ we used the partial derivatives
for the Johnson $V$ band computed by \citet{Kowalczuk}.  The value of 
$D_{1,l}$ is typically around 40 and $D_{3,l}$ is in the range 2--3.  
The geometric term, $D_{2,l}$ increases in absolute value with $l$ and 
is comparable with $D_{1,l}$ for high $l$.  The temperature variation 
together with $b_l$ dominates the visibility.

\section{The effect of rotation}

In a non-rotating star, pulsation modes with the same spherical harmonic
degree, $l$, but with different azimuthal numbers, $m$, have the same frequency.
To first order, the effect of rotation is to remove this degeneracy so that
a mode with frequency $\nu_0$ in the non-rotating star appears as $2l + 1$
equally-spaced multiplets with frequencies, $\nu$, given by
\begin{align*}
\nu &= \nu_0 + m\nu_{\rm rot}\left(1 - C_{nl}\right),
\end{align*}
where $\nu_{\rm rot}$ is the rotation frequency and $C_{nl}$ is a constant
which depends on the structure of the star, $l$ and the radial order, $n$
\citep{Ledoux1951}.  The value of $C_{nl}$ is quite small for p modes and
hence the frequency splitting may be several cycles per day in the extreme
case of high $m$ and large $\nu_{\rm rot}$.  The effect of rotation to first
order is to introduce a symmetric spread in frequencies.

The above formula breaks down as the rotation rate increases and the splitting
is no longer symmetric.  The frequency of each multiplet decreases with
increasing rotation rate, but the decrease in frequency is greater for
larger values of $|m|$.   Thus the spread in frequencies is skewed towards
low frequencies.  It is therefore important to take this effect into account
by including higher-order rotational perturbations.

The following formula, \citep{Goupil2011}, is accurate to third order:
\begin{align*}
\omega &= \omega_0 - m(1  - C_{nl})\Omega + (D_1 + m^2D_2)\Omega^2
+ m(T_1 + m^2T_2)\Omega^3,
\end{align*}
where $\omega$, $\omega_0$ are the perturbed and unperturbed angular
pulsation frequencies and $\Omega$ is the angular rotation frequency.  The
values of $D_1$, $D_2$, $T_1$, and $T_2$ depend on the mode.  In this 
expression the pulsation and rotation frequencies are in units of the dynamical
frequency $\omega_{\rm dyn} = \sqrt{GM/R^3}$ where $G$ is the gravitational constant,
$M$ the stellar mass and $R$ the stellar polar radius.  The coefficients have
not been calculated for stars in the $\delta$~Sct instability strip.  However,
\citet{Reese2006} lists values for a polytrope of index 3 for $l \le 3$ and
$n \le 10$.

Since most A stars are moderate or rapid rotators, it is very important to
include rotation effects as accurately as possible.  Since calculations are
only available for polytropes and for limited values of $l$ and $n$, we are
faced with a problem.  We decided that the best approach is to use the
third-order formula and polytropic values for all p modes.  For p modes with
$l > 3$ we use the coefficients for $l = 3$.  For radial orders greater
$n > 10$ we use the values for $n = 10$. This might overestimate the rotational
splitting for modes with $l > 3$.  Rotational splitting for g modes is smaller
than for p modes.  We decided to use the first-order formula for all g modes.
The distinction between p and g modes is made on the basis of the ratio of
the mode kinetic energy in the gravity-wave propagation zone, $E_{kg}$, 
to the total kinetic energy of a mode, $E_k$, as calculated by the models.  
We treated modes with $E_{kg}/E_k < 0.5$ as p modes.

\begin{figure}
\centering
\includegraphics{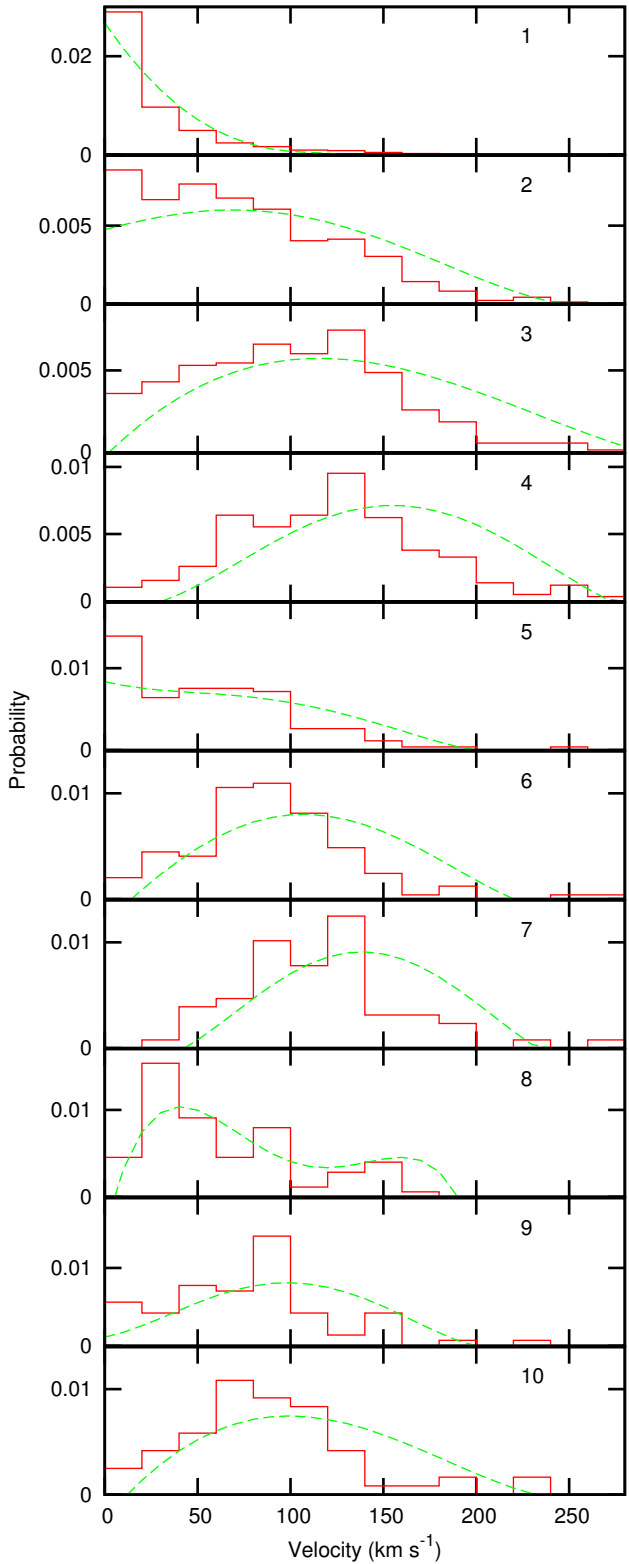}
\caption{The distribution of projected rotation velocity, $v \sin i$, for
field stars with spectral types appropriate for each region is shown by the
histogram.  The distribution of true equatorial rotational velocities, $v_e$, 
is shown by the dashed curve.}
\label{vdist}
\end{figure}

To calculate the frequencies of the rotational multiplets in the way just
described requires knowledge of the rotation frequency, $\nu_{\rm rot}$.
In \citet{Balona2014a} the value of $\nu_{\rm rot}$ was obtained from a peak
in the periodogram which could be attributed to rotational modulation.
We cannot do this because the number of stars with known photometric
rotation period is too small (and sometimes zero) in each region.
To overcome this problem we have to make the assumption that the
distribution of equatorial rotational velocities in a particular region
closely corresponds to the distribution of equatorial rotational velocities
for field stars with a spectral type and luminosity class appropriate to
that region. We used the catalogue of projected rotational velocities
compiled by \citet{Glebocki2000} to determine the distribution of $v \sin i$
for stars in each region.  To obtain the distribution of true equatorial
rotational velocities, we used the procedure described in \citet{Balona1975}.
Fig.\,\ref{vdist} shows the distribution of $v \sin i$ and the polynomial
approximation to the true distribution of equatorial rotational velocities.

\section{Assigning weights}

For a stellar model with a given mass and radius we may calculate the
rotational frequency, $\nu_{\rm rot}$, for any given equatorial rotational
velocity, $v_e$.  Given $\nu_{\rm rot}$ and the calculated frequency of a
unstable mode of degree $l$ and radial order $n$, we may calculate the
$2l+1$ rotationally-perturbed frequencies using the first-order Ledoux
formula for g modes and the third-order formula for the p modes.  To each
of these frequency multiplets we assign the same weight, $w_{\rm rot}$,
which is the probability of finding a star with the particular value of
$v_e$.  This weight is calculated using the polynomial fit to the distribution
of true equatorial velocities for stars in the appropriate region.  We
calculated the frequencies and weights of all rotationally-split multiplets
using values of $v_e$ from zero to the maximum value in steps of 10\,km\,s$^{-1}$.

We also assign a weight $w_i = \langle Y_l^m (i, 0)\rangle$ which accounts for
the visibility of the mode due to random orientation of the axis of rotation
by numerically calculating the integral.  This term is relatively
unimportant because the $w_i$ does not change much.  The visibility due to
cancellation effects, $w_b = b_l$, is pre-calculated for any given value of
$l$.  Finally we calculate the weight $w_f = D_{1,l} + D_{2,l} + D_{3,l}$
using the values of $f, \omega$ and $\bar{\rho}$ from the model as well as
the partial derivatives applicable to the given model.  The total weight for
the mode with frequency $\nu$ is $W_\nu = w_{\rm rot}w_iw_bw_f$.
In calculating the predicted frequency distribution we chose a bin size of
1\,d$^{-1}$.  The probability at any given frequency, $P(\nu)$, is $P(\nu) =
\sum W_\nu$ for all frequencies within this interval.

In principle, the frequency distribution, $P(\nu)$, needs to be calculated for
all possible values of $l$.  However, the mode visibility drops sharply with
$l$, so that modes with $l > l_{\rm max}$ have amplitudes which are so low
that they can be ignored.  The simplest way to determine $l_{\rm max}$ is to
construct frequency distributions using increasing values of $l_{\rm max}$.
We find that there is scarcely any difference in the distributions with
$l_{\rm max} > 4$.  We adopted $l_{\rm max} = 6$ which is certainly adequate
for our purposes.

In the above procedure we have made assumptions that may not perhaps be
entirely justifiable.  For example, we assume that all rotationally-split
multiplets have the same intrinsic amplitude.  We cannot calculate
these amplitudes so we do not really know, but this is probably a fair
assumption when averaged over many stars.  It seem reasonable to
assume that stars with approximately the same stellar parameters and
rotational velocities have similar frequency distributions.  This is
certainly the case given our current understanding and that is what the
models predict.  We shall see below that even this seemingly secure
assumption may not be correct.

\begin{figure*}
\centering
\includegraphics{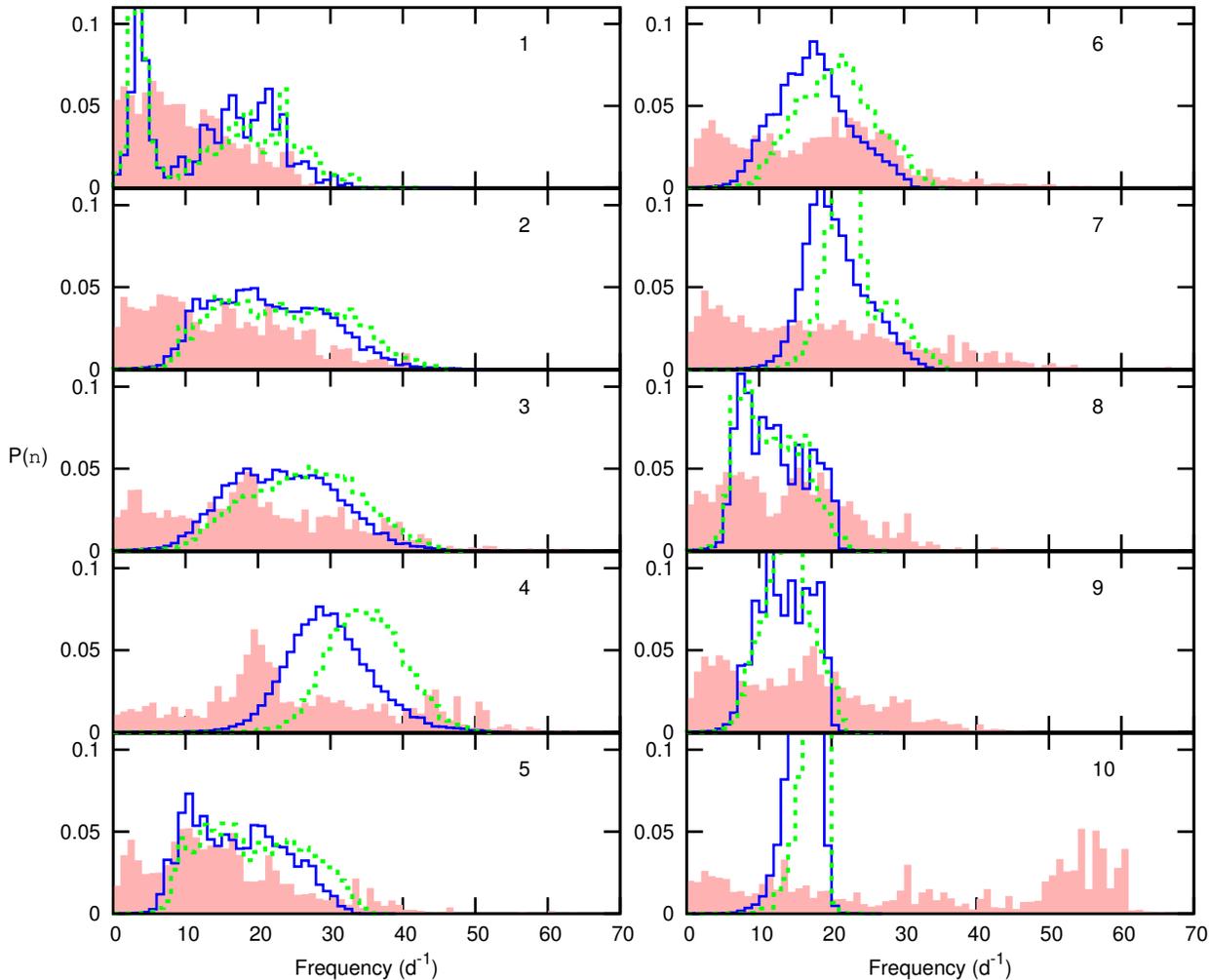}
\caption{The observed frequency distributions (filled) and the corresponding
theoretical frequency distributions derived from pulsation models with
metal abundance $Z = 0.015$ (solid blue curves) and $Z = 0.030$ (dotted
green curves).  Only unstable modes with the degrees, $l \le 6$ are considered.
Each panel shows the frequency distribution corresponding to
the labeled region in the H-R diagram of Fig.\,\ref{regions}.}
\label{dist}
\end{figure*}

\begin{figure*}
\centering
\includegraphics{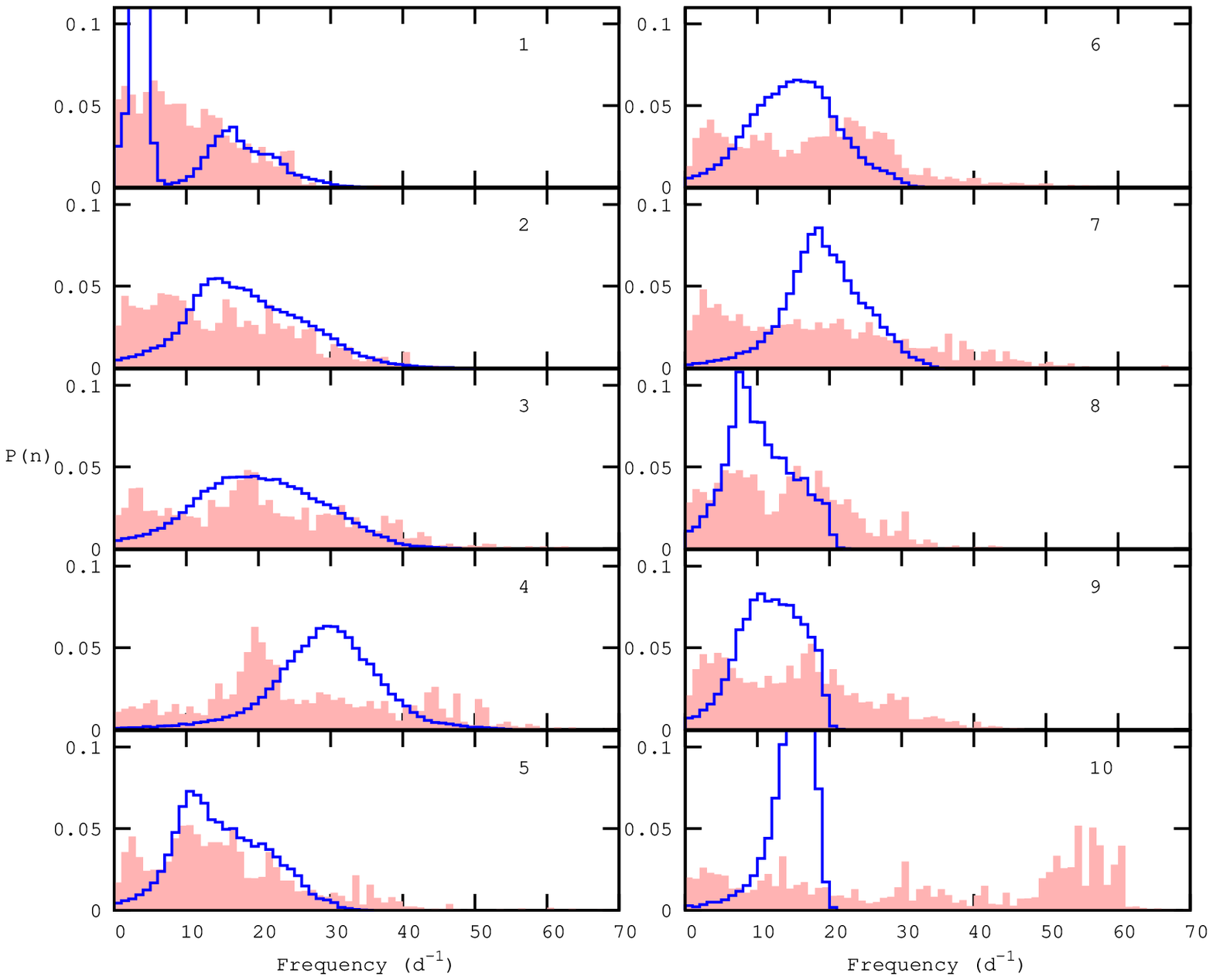}
\caption{The same as Fig.\,\ref{dist}, but we have artificially increased the
effect of rotational splitting by using the 3-rd order formula for all modes
and setting the disc-averaging factor, $b_l$, to unity so as to maximize the
visibility of modes of high degree where rotational splitting is greatest.
Only predicted distributions for $Z = 0.015$ are shown.}
\label{distbl}
\end{figure*}

\section{Comparison with theoretical predictions}

Equilibrium models were computed using the Warsaw - New Jersey evolution code
\citep{Paczynski1970}, assuming two initial hydrogen fractions, $X_0 = 0.70$,
$0.65$, and four metal abundances, $Z=0.005$, 0.015, 0.030 and 0.050.  We 
adopted the chemical element mixture of \citet{Asplund2009} and two sources of 
the opacity data: OPAL \citep{Rogers1992} and OP \citep{Seaton2005}.  We did 
not include overshooting from the convective core and used the mixing length 
parameter $\alpha_{\rm MLT} = 0.5$ for the convective scale height.  We chose 
this low value of $\alpha_{\rm MLT}$ to eliminate, as far as possible, the 
fictitious instability of low-frequency modes in the cooler and lower-mass models.  
This instability is a consequence of the frozen convective flux approximation 
that we adopted.

We calculated a grid of equilibrium models covering the $\delta$ Sct instability 
strip.  The models take rotation into account to first order by applying the 
centrifugal force correction to local gravity while keeping spherical symmetry.  
We used a typical initial equatorial rotational velocity of $v_e = 150$\,km\,s$^{-1}$ 
except for the models with the lowest mass where we used $v_e = 50$\,km\,s$^{-1}$.
The effect of rotation is a slight shift of the evolutionary tracks to lower 
effective temperatures and higher luminosities.  The main sequence band is 
also somewhat extended (see, for example, \citealt{Breger1998b}).  Given the 
fact that we need to determine the rotational frequency splitting over a wide 
range of rotational velocities corresponding to the known equatorial rotational 
velocity distribution for a particular group of stars, the value of $v_e$ in the 
equilibrium model is not an important factor.  A non-rotating model would serve 
our purposes equally well.  As explained in the previous section, we used the 
3-rd order rotational perturbation only for p modes, while for g modes the 
1-st order Ledoux splitting formula was used.  

The linear nonadiabatic code of \citet{Dziembowski1977a} was used to determine 
the frequencies and the instability parameter, $\eta$, as well as the the ratio 
of luminous flux to displacement, $f$, for each mode.  Only modes with 
$l \le 6$ were considered.  

In calculating the predicted frequency distribution, we use individual
values of $\log T_{\rm eff}$ and $\log L/L_\odot$ for each star in a particular
region.  The model with temperature and luminosity closest to these values
is used to determine the predicted frequency distribution.  The average of all
frequency distributions for stars in a given region is taken as the best
approximation of the predicted frequency distribution for that region.

The observed and predicted frequency distributions are shown in
Fig.\,\ref{dist} for $Z = 0.015$ and $Z = 0.030$ using OPAL opacities and
an initial hydrogen abundance $X_0 = 0.70$.  The unstable low frequencies 
predicted for the models of region 1 are a result of the frozen convection 
approximation and are fictitious.  It can be seen that low frequencies are 
observed in stars across the whole instability strip.  In other words, all 
$\delta$~Sct stars are hybrids.  Also, it can be seen that the predicted 
frequency distributions fail to show frequencies less than about 5\,d$^{-1}$. 
The same result was obtained with models calculated using OP opacities.

Apart from the disagreement in the low-frequency range, we also note that
for the more luminous stars of regions 8, 9 and 10 the observed distributions
extend to much higher frequencies than predicted.  In general, we expect the
observed distribution to be broader than the predicted distribution because
each region includes a fraction of stars with true values of $T_{\rm eff}$
and $\log L/L_\odot$ which are outside the region.  Even so, it does not
seem possible to explain the high-frequency tails of the luminous stars in
this way because there is no such tail among the less luminous stars.

What we can definitely conclude from Fig.\,\ref{dist} is that a change in
metal abundance cannot explain the mismatch between observations and the
models.  We also calculated models with $Z = 0.005$ and $Z = 0.050$ which 
we do not show in Fig.\,\ref{dist} to avoid confusion.  The distributions 
with lower and higher metal abundance are quite similar to those shown in the 
figure.  We also calculated models where the helium abundance is substantially 
increased so that the initial hydrogen abundance is decreased from  $X_0 = 0.70$ 
to $X_0 = 0.65$, but this does not affect mode stability at all at low 
frequencies.  However, there is a slight increase of the instability 
parameter, $\eta$, at high frequencies.  We also studied the effect of modes
of higher degree and found a negligible difference in the distributions 
for modes with $l_{\rm max} = 6$ and $l_{\rm max} = 10$.

It could be argued that we have not taken the effect of rotation fully into
account and that agreement could be achieved by a more realistic treatment
of rotational splitting.  Unfortunately, we do not have access to
non-adiabatic models which take into account stellar distortion, gravity
darkening and rapid rotation in a more realistic way.  We can, however, take
the effect of rotation to extremes by using the 3-rd order perturbation for
all modes.  We can also enhance the visibility of modes of high degree
(which have maximum rotational splitting) by artificially suppressing the
disc-averaging factor, $b_l$ (i.e, setting $b_l = 1$ for all modes).
The distributions calculated in this way are shown in Fig.\,\ref{distbl} for
$Z = 0.015$.

Although there is a slight increase in the numbers of low-frequency modes,
it is still far too little to match observations.  Artificially increasing
the effect of rotation in this way also does not assist in explaining the
high-frequency tail in luminous stars.  We may confidently conclude that
there is no possibility that rotational splitting of high-frequency modes can
can explain the low frequencies.

\begin{figure}
\centering
\includegraphics{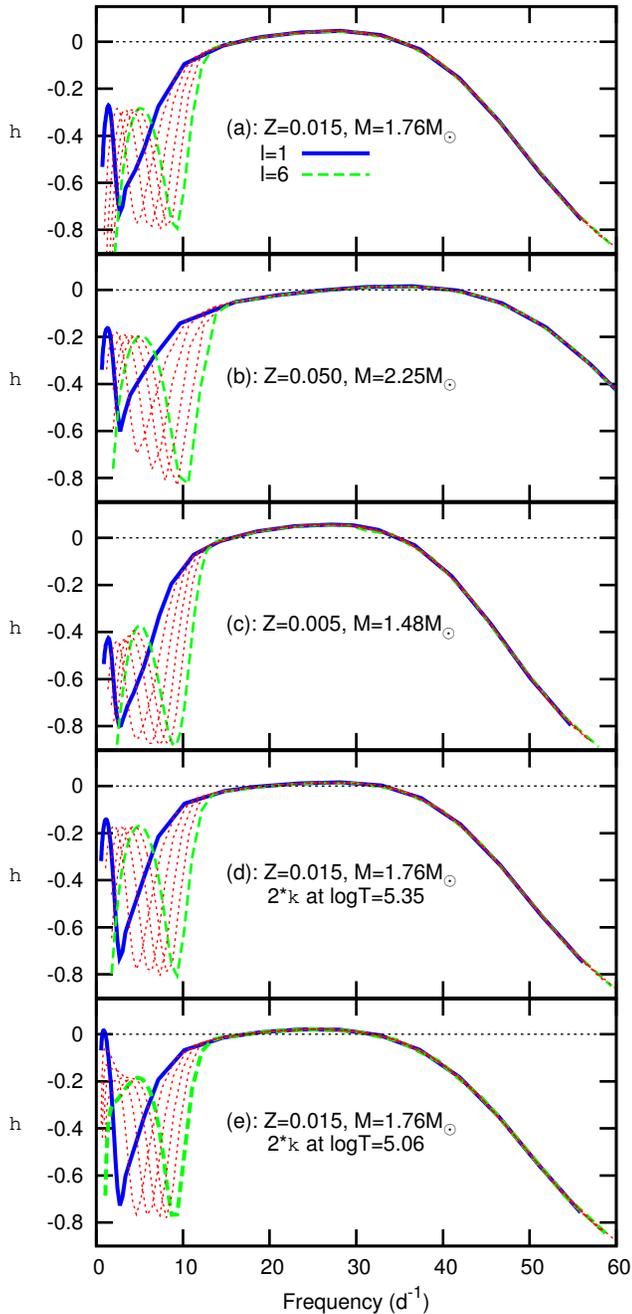}
\caption{The instability parameter, $\eta$, as a function of frequency for modes
with $l \le 6$ for representative models of region 3. All models have
$\log T_{\rm eff} = 3.8855$ and $\log L/L_\odot = 1.093$.  Panel (a) is
a standard model with mass $M = 1.76$\,$M_\odot$ and $Z = 0.015$.  Panel (b)
is a standard model with $M = 2.25$\,$M_\odot$ and $Z = 0.050$. Panel (c)
is a standard model with $M = 1.48$\,$M_\odot$ and $Z = 0.005$.  In panel
(d) the opacity at $\log T = 5.35$ (the Z-bump region) has been increased by
a factor of two.  In panel (e) the opacity at $\log T = 5.06$ has been 
increased by a factor of two to simulate a new opacity bump in the Kurucz 
opacity data (see main text).  All models have initial hydrogen abundance 
$X_0 = 0.70$.}
\label{eta}
\end{figure}

Since the low frequencies cannot be explained by modifications of standard
models, we need to consider other possibilities.

Low frequencies not predicted by models also occur in the $\beta$~Cep stars. 
These are early B stars with multiple p, g and mixed modes driven by the 
opacity bump due to iron group elements (the Z bump).  Some low frequencies in 
these stars can be explained by standard models and some demand increasing the 
opacity in the Z-bump region \citep{Pamyatnykh2004}.  Whether this is the correct 
solution or how to create the increased opacity is not known.  As in the B stars, 
it is possible that artificially increasing the opacity of $\delta$~Sct models 
in the Z-bump region at $\log T \approx 5.35$ might destabilize the low 
frequencies. 

Recently, \citet{Cugier2012, Cugier2014} found that the OPAL and OP 
opacities are markedly underestimated in comparison with the Rosseland mean 
opacities taken from the \citet{Castelli2003} model atmospheres.  As a
result, a new opacity bump appears at $\log T \approx 5.06$.  This bump is due 
to an attempt to include all spectral lines of atoms in the opacities \citep{Kurucz2011}.
This additional opacity bump  affects the stability of stellar pulsations and 
also needs to be considered.  In testing the effect of this bump (which we
will call the ``Kurucz'' bump), we did not use the actual Kurucz opacities
where the bump occurs, but simulate this additional opacity bump by artificially 
increasing the standard OPAL opacities in the appropriate temperature range.

In Fig.\,\ref{eta} we show the instability parameter, $\eta$, as a function
of frequency for $l \le 6$ in five models with the same effective temperature 
and luminosity but different metal abundances and enhanced opacities.
As one can see, an increase in $\eta$ in the low frequency range can be
obtained either by increasing the value of Z or by an increased opacity
in the Z-bump or by increasing the standard OPAL opacities to simulate the 
Kurucz bump.  However, low-frequency modes remain stable except when the 
Kurucz bump opacity is included.  Instability of $l=1$ modes at low frequencies 
begins when the standard OPAL opacity in the Kurucz-bump region is increased 
by a factor of two.  Maximum instability occurs when this opacity is increased 
by a factor of three.  Any further opacity increase results in saturation and 
no further increase in $\eta$ can occur.  Note that an increase in opacity in 
the $Z-$bump or Kurucz-bump regions leads to somewhat reduced $\eta$ for modes 
with high frequencies.   For example, the frequency range of unstable 
high-frequency modes is decreased from the 15--35\,d$^{-1}$ range for a standard 
model to 16--32\,d$^{-1}$ for a model where the OPAL opacity in the Kurucz bump 
region is increased by a factor of two.

\begin{figure}
\centering
\includegraphics{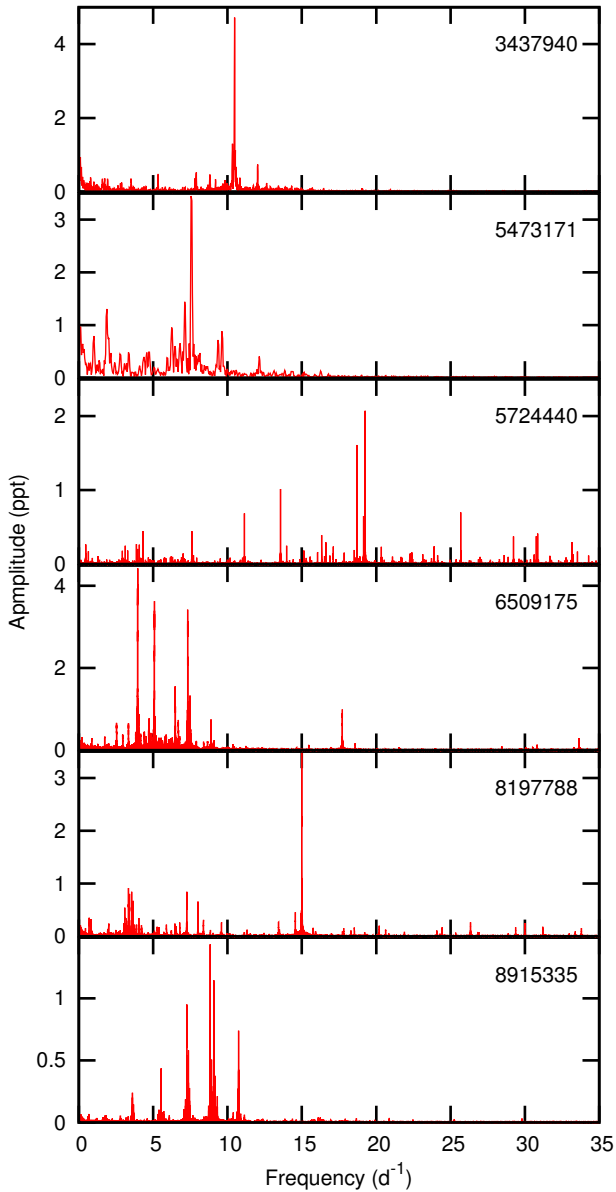}
\caption{Examples of periodograms of $\delta$~Sct stars with similar
effective temperatures and luminosities (in region 3).  All stars (KIC
numbers are shown) have spectroscopic determinations of effective
temperature and surface gravity.}
\label{morph03}
\end{figure}

\section{Further unsolved problems}

Even a cursory inspection of the periodograms of $\delta$~Sct stars is
sufficient to show the wide variety of frequency patterns in these stars.
In fact, each star is unique and can be identified by its periodogram.
This is strange because one expects stars with similar effective
temperatures, luminosities and rotational velocities to have similar
frequency spectra, but this does not seem to be the case.  Examples of this
effect are shown in Fig.\,\ref{morph03} for stars belonging to region 3.
The stars shown all have well-determined spectroscopic effective 
temperatures and surface gravities.

While it is true that the stars may individually have considerably different
parameters due to substantial errors in $T_{\rm eff}$ and $\log L/L_\odot$,
the differences in frequency spectra are very pronounced.  This disparity
cannot be proved until more accurate parameters are obtained, of course.
Nevertheless, it seems that small differences in the parameters may lead to
large differences in frequency patterns, implying that nonlinearities in
$\delta$~Sct envelopes may be very important.

The question of the number of non-pulsating stars in the $\delta$~Sct
instability strip was addressed by \citet{Balona2011g}.  They found that
most $\delta$~Sct stars have effective temperatures in the range $7000 <
T_{\rm eff} < 8500$\,K and that even in this range no more than 40--50 per
cent of stars pulsate as $\delta$~Sct variables.  We can confirm this
finding.  We have found 1165 $\delta$~Sct stars in this temperature range in
the {\it Kepler} field observed in long-cadence mode.  There are 2839 stars
(including the $\delta$~Sct stars) in the same temperature range, meaning
that only about 41\,per cent of stars in the instability strip are detected
as $\delta$~Sct stars in the {\it Kepler} photometry.

It could be argued that most of the non-$\delta$~Sct stars are outside the
instability strip due to errors in the effective temperature.  Errors in
luminosity alone play no role because the stars will still be within the
instability strip as they can only lie between the ZAMS and TAMS.  There are
no known high-luminosity stars in this temperature range and in any case none
are expected due to the very short lifetimes such stars will have in this
evolutionary stage.  We have seen that a typical error of about 200\,K in
$T_{\rm eff}$ can be expected, which is considerably less than the range of
1500\,K that we are discussing.  The probability that a star in the middle
of the instability strip lies, in actual fact, outside the strip is less
than 0.01.  Of course, the probability will be higher if the star is closer
to the edge of the instability strip, but it means that the probability
that all 1674 non-pulsating stars are outside the instability strip is
the product of the individual probabilities which is essentially zero.  
There can be no question that non-pulsating stars are present in the 
$\delta$~Sct instability strip.  Of course, it can be argued that the 
pulsations are less than the detectable limit for {\it Kepler} (typically 
less than about 50\,ppm).  This might be the case, though one can argue 
against this idea on the basis of the amplitude distribution 
\citep{Balona2011g}.  Such an amplitude disparity, if it exists, 
still needs an explanation.

\section{Conclusions}

We have compared the frequency distributions of groups of $\delta$~Sct stars
with similar effective temperatures and luminosities with the frequency
distributions calculated from pulsation models taking into account, in an
approximate but fairly realistic way, the effect of rotational splitting.
We have come to the same conclusion as \citet{Balona2014a}: that it is impossible to
account for frequencies with $\nu < 5$\,d$^{-1}$ on the basis that these are
high-frequency modes shifted to low frequencies by rotation.  Other factors
examined in detail in \citet{Balona2014a} are also excluded.  The presence
of low frequencies is a general feature of all $\delta$~Sct stars, no matter
where they lie in the instability strip.  This is clearly seen in the
observed distributions shown in Fig.\,\ref{dist}. We also found other
inconsistencies between the observed and predicted frequency distributions.
For example, the observed distribution for the more luminous stars is much
wider than expected.  However, this is a minor problem compared to absence
of predicted low frequencies.

We calculated a large variety of standard models with differing helium and
metal abundances, but in no case could we find unstable modes at low
frequencies. The instability parameter, $\eta$, does however tend to
increase as the metal abundance, Z, increases.  We also studied the effect 
of increasing the opacity at temperatures $\log T =5.35$ (the Z bump) and the 
effect of simulating a new opacity bump at $\log T =5.06$ (the Kurucz bump).  
The Kurucz bump does not occur in the OPAL and OP opacities, but appears
in model atmospheres, as discussed by \citet{Cugier2012, Cugier2014}.  We 
found that increasing the opacity in these two regions does increase the 
value of $\eta$.  However, low-frequency modes remain stable unless the
OPAL opacities at $\log T = 5.06$ are increased by at least a factor of two
to simulate the Kurucz bump or if the opacity in the Z bump is increased 
by a factor of at least three.  At the same time, the range of high-frequency 
modes is decreased to some extent.  We do not claim that this opacity increase 
is the solution to the problem of low frequencies in $\delta$~Scuti stars, but 
only that there is a likely problem with current opacity data and that further 
investigation into the sources of opacity in this region are required.

A problem of equal importance is the question of why there are so many
non-pulsating stars in the $\delta$~Sct instability strip.  This could
simply be that the {\it Kepler} photometry is not sufficiently precise
to detect such pulsations.  This does not resolve the problem because it
would not explain the very large disparity in amplitudes among the
$\delta$~Sct stars.  We are clearly faced with substantial problems in 
the physics of A--F stars.

Recently, further problems in our current understanding of stellar
pulsations have come to light.  It seems that there are a group of pulsating
stars with high frequencies lying between the cool end of the $\beta$~Cep
instability strip and the hot end of the $\delta$~Sct instability strip.
These were recently investigated by \citet{Balona2015c} using {\it Kepler}
and K2 data.  Standard models cannot reproduce these pulsations, though
it is possible that these so-called ``Maia'' variables may be
rapidly-rotating SPB stars.  The unexplained presence of low
frequencies in a well-studied group such as the $\delta$~Sct stars shows
that we still do not fully understand the physics and/or envelopes of hot
stars.  Until we have a better understanding of the low frequencies in A
stars we are not likely to make much progress with the more complex problem
of explaining the origin of high frequencies in Maia variables.

\section*{Acknowledgments}

The authors wish to thank the {\it Kepler} team for their generosity in allowing
the data to be released to the Kepler Asteroseismic Science Consortium (KASC)
ahead of public release and for their outstanding efforts which have made these
results possible. Funding for the {\it Kepler} mission is provided by NASA's
Science Mission Directorate.

LAB wishes to thank the National Research Foundation of South Africa for financial
support. JDD and AAP acknowledge partial financial support from the Polish NCN grants 
2011/01/B/ST9/05448 and 2011/01/M/ST9/05914.

\bibliographystyle{mn2e}
\bibliography{model}

\label{lastpage}

\end{document}